# Mutual Conversion Between Preference Maps and Cook-Seiford Vectors

Fujun Hou

School of Management and Economics, Beijing Institute of Technology, Beijing, China, 100081,
e-mail: houfj@bit.edu.cn

**Abstract:** In group decision making, the preference map and Cook-Seiford vector are two concepts as ways of describing ties-permitted ordinal rankings. This paper shows that they are equivalent for representing ties-permitted ordinal rankings. Transformation formulas from one to the other are given and the inherent consistency of the mutual conversion is discussed. The proposed methods are illustrated by some examples. Some possible future applications of the proposed formulas are also pointed out.

## 1. Introduction

In decision making and group decision making (for short, GDM), there are essentially three different forms of ranking orders for people to express preference rankings: the linear order, the weak order and the partial order (Cook, et al., 1986). A linear ordering corresponds to a complete ranking of alternatives. In this case, alternatives are well differentiated and ordered hence no ties included. A weak ordering corresponds to a situation that alternatives are all compared and ties are permitted. In a partial ordering, there may be such pairs of alternatives that are not compared, but ties are allowed. Therefore, experts can use ties-permitted ordinal rankings to express preferences when their preferences are constructed based upon weak order relations. This paper is concerned with two concepts as ways of describing ties-permitted ordinal rankings, particularly, the Cook-Seiford vector and the preference map as well as their mutual conversion.

The paper is organized as follows. Some preliminaries are introduced in Section 2. Section 3 contains our main results including the transformation formulas and the inherent consistency of the mutual conversion. The proposed methods are illustrated by some examples. Section 4 concludes the paper and points out some possible future applications of the proposed formulas.

## 2. Preliminaries

Suppose that there are an alternatives set, denoted $\{x_1, x_2, \ldots, x_n\}$, and an experts set, denoted $\{E_1, E_2, \ldots, E_m\}$, where $1 < m, n < +\infty$. In what follows we assume that the experts' preferences are provided by ties-permitted ordinal rankings. In other words, experts give their preferences based on weak order relations on the alternatives set.

### 2.1. Cook-Seiford vector

In Cook-Seiford GDM approach, a ties-permitted ordinal ranking is represented by a Cook-Seiford vector (for short, C-S vector), whose elements indicate alternatives' ranking positions (Armstrong et al., 1982; Cook and Seiford, 1978; Cook, 2006; Cook et al., 1996). If several alternatives are ranked in a tie, then these alternatives will all be assigned to the middle position of the tie (Cook and Seiford, 1978). For instance, let $x_1 \succ x_2 \sim x_3 \succ x_4$ be a ties-permitted ordinal ranking. Then its corresponding C-S vector is

$$\begin{matrix} x_1 \\ x_2 \\ x_3 \\ x_4 \end{matrix} \begin{bmatrix} 1 \\ 2.5 \\ 2.5 \\ 4 \end{bmatrix},$$

which indicates that alternative $x_1$ should be ranked at position 1, alternatives $x_2$ and $x_3$ should both be assigned to position 2.5 since they are in a tie and will occupy positions 2-3, and $x_4$ should be ranked at position 4.

## 2.2. Preference map

Preference map (for short, PM) is preference sequence for representing a ties-permitted ordinal ranking. It is a 'vector' whose entries are sets which contain alternatives' possible ranking positions. This concept was first introduced and named as 'preference sequence vector' by Hou in 2015 (Hou, 2015a; Hou, 2015b), and later renamed as 'preference map' by Hou and Triantaphyllou (Hou and Triantaphyllou, 2018).

**Definition 1.** A sequence $[PM_i]_{n \times 1}$ is defined as a preference map (PM) corresponding to a ties-permitted ordinal ranking on the alternative set $X = \{x_1, x_2, \ldots, x_n\}$ with respect to a weak order relation $\preccurlyeq$ such that

$$PM_i = \{|P_i|+1, |P_i|+2, \ldots, |P_i|+|S_i|\}, \tag{1}$$

where $P_i = \{x_j \mid x_j \in X, x_j \succ x_i\}$ is alternative $x_i$'s dominated set that contains the alternatives predominating $x_i$; and $S_i = \{x_j \mid x_j \in X, x_j \sim x_i\}$ is alternative $x_i$'s indifference set that contains the alternatives similar to $x_i$.

For sake of convenience, we reconsider the ties-permitted ordinal ranking in subsection 2.1, that is, $x_1 \succ x_2 \sim x_3 \succ x_4$. We have

- Regarding $x_1$, its dominated set and indifference set are $P_1 = \varnothing$ and $S_1 = \{x_1\}$, respectively;
- Regarding $x_2$, its dominated set and indifference set are $P_2 = \{x_1\}$ and

$S_2 = \{x_2, x_3\}$, respectively;

- Regarding $x_3$, its dominated set and indifference set are $P_3 = \{x_1\}$ and $S_3 = \{x_2, x_3\}$, respectively;

- Regarding $x_4$, its dominated set and indifference set are $P_4 = \{x_1, x_2, x_3\}$ and $S_4 = \{x_4\}$, respectively.

According to formula (1), the PM corresponding to $x_1 \succ x_2 \sim x_3 \succ x_4$ is

$$\begin{array}{c} x_1 \\ x_2 \\ x_3 \\ x_4 \end{array} \begin{bmatrix} \{1\} \\ \{2,3\} \\ \{2,3\} \\ \{4\} \end{bmatrix}.$$

Because a ties-permitted ordinal ranking is constructed based on a weak order relation $\preccurlyeq$ on the alternative set, thus a ties-permitted ordinal ranking implies an equivalence relation on the alternative set with respect to the indifference relation $\sim$. Consequently, a PM also indicates an equivalence relation on the alternative set. Because the entries of a PM represent the alternatives' possible ranking positions, thus different entries of a PM constitute of a partition of the set $\{1, 2, \ldots, n\}$. In other words, an entry of a PM is a partition block of set $\{1, 2, \ldots, n\}$ (Hou, 2015a,b).

## 3. Main results

### 3.1. From preference map to Cook-Seiford vector

**Definition 2.** Let $PM = [PM_i]_{n \times 1}$ be a preference map. Its corresponding C-S vector is defined by $CS = [CS_i]_{n \times 1}$, where

$$CS_i = \frac{\max PM_i + \min PM_i}{2}. \tag{2}$$

To illustrate, we reconsider the PM that appeared in Section 2 and another in the following:

$$PM^{(1)} = \begin{bmatrix} \{1\} \\ \{2,3\} \\ \{2,3\} \\ \{4\} \end{bmatrix}, \text{ and } PM^{(2)} = \begin{bmatrix} \{1,2\} \\ \{1,2\} \\ \{3\} \\ \{4\} \end{bmatrix}.$$

According to formula (2), their corresponding C-S vectors are

$$CS^{(1)} = \begin{bmatrix} 1 \\ 2.5 \\ 2.5 \\ 4 \end{bmatrix}, \text{ and } CS^{(2)} = \begin{bmatrix} 1.5 \\ 1.5 \\ 3 \\ 4 \end{bmatrix},$$

respectively.

### 3.2. From Cook-Seiford vector to preference map

**Definition 3.** Let $CS = [CS_i]_{n \times 1}$ be a C-S vector. Its corresponding preference map is defined by $PM = [PM_i]_{n \times 1}$, where

$$PM_i = \{a_i, a_i + 1, a_i + 2, \ldots, b_i\} \tag{3}$$

with

$$\begin{cases} a_i = c_i - \frac{1}{2}(d_i - 1) \\ b_i = c_i + \frac{1}{2}(d_i - 1) \\ c_i = CS_i \\ d_i = \sum_{j=1}^{n} \delta_{ij}, \ \delta_{ij} = \begin{cases} 0, & CS_j \neq CS_i \\ 1, & CS_j = CS_i \end{cases} \end{cases}, \tag{4}$$

where $c_i$ represents the center of $PM_i$, $d_i$ the number of alternatives that are similar to alternative $x_i$, $a_i$ the minimum value among the elements of $PM_i$, and $d_i$ the maximum value among the elements of $PM_i$.

For illustration, the C-S vectors of $CS^{(1)}$ and $CS^{(2)}$ in subsection 3.1 are reconsidered. The computation results, which are computed by using formulas (3) and (4), are summarized in Table 1.

**Table 1.** Computation results of converting C-S vectors to PMs

| C-S vector | $CS_i$ | $c_i$ | $d_i$ | $a_i$ | $b_i$ | Corresponding PMs |
|---|---|---|---|---|---|---|
| $CS^{(1)} = \begin{bmatrix} 1 \\ 2.5 \\ 2.5 \\ 4 \end{bmatrix}$ | 1<br>2.5<br>2.5<br>4 | 1<br>2.5<br>2.5<br>4 | 1<br>2<br>2<br>1 | 1<br>2<br>2<br>4 | 1<br>3<br>3<br>4 | $PM^{(1)} = \begin{bmatrix} \{1\} \\ \{2,3\} \\ \{2,3\} \\ \{4\} \end{bmatrix}$ |
| $CS^{(2)} = \begin{bmatrix} 1.5 \\ 1.5 \\ 3 \\ 4 \end{bmatrix}$ | 1.5<br>1.5<br>3<br>4 | 1.5<br>1.5<br>3<br>4 | 2<br>2<br>1<br>1 | 1<br>1<br>3<br>4 | 2<br>2<br>3<br>4 | $PM^{(2)} = \begin{bmatrix} \{1,2\} \\ \{1,2\} \\ \{3\} \\ \{4\} \end{bmatrix}$ |

### 3.3. Inherent consistency of the mutual conversion

As can be seen from Table 1 and those computation results in subsection 3.1, if one converts a PM to a C-S vector, and then converts that C-S vector to a PM, then the second PM vector will always be equivalent to the first PM. This is true as a result of the fact that: (1) A ties-permitted ordinal ranking, which is constructed based on a weak order relation on the alternative set, represents a unique ordered partition of the alternative set; and (2) C-S vector and the PM are both efficient means for describing ties-permitted ordinal rankings.

For purpose of intuition, we depict the above analysis by Figure 1 in a particular case where the ties-permitted ordinal ranking is $x_1 \succ x_2 \sim x_3 \succ x_4$.

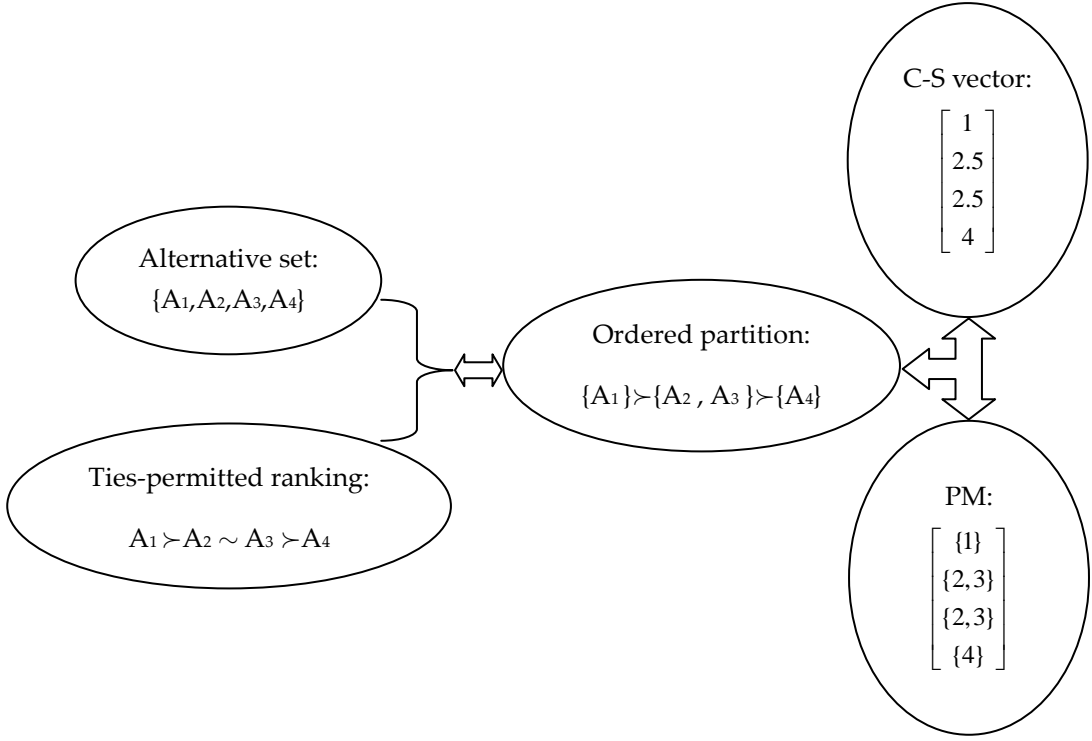

**Figure 1**. Relationship of preference ranking, C-S vector and PM

Based on the understanding of Figure 1, we have the following result.

**Theorem 1**. Let $PM = [PM_i]_{n\times 1}$ and $CS = [CS_i]_{n\times 1}$ be a PM and a C-S vector, respectively. Suppose that $PM_i = \{pm_{i1}, pm_{i2}, \ldots, pm_{iL_i}\}$ and $sum(PM_i) = \sum_{j=1}^{L_i} pm_{ij}$. We have

$$\sum_{i=1}^{n} CS_i = \sum_{i=1}^{n} \frac{sum(PM_i)}{|PM_i|} = \frac{1}{2}n(n+1).$$

**Proof**. As mentioned in Section 2, an entry of a PM corresponds to a partition block of $\{1, 2, \ldots, n\}$ which comes from an equivalence relation that is constructed on the alternative set. Therefore, we have $\sum_{i=1}^{n} \frac{sum(PM_i)}{|PM_i|} = (1 + 2 + \ldots + n) = \frac{1}{2}n(n+1)$.

From formulas (2) and (3), we know that $CS_i = \frac{1}{2}(a_i + b_i)$. Because $sum(PM_i) = a_i + (a_i + 1) + (a_i + 2) + \ldots + b_i = \frac{1}{2}|PM_i|(a_i + b_i)$, thus we have $CS_i = \frac{1}{2}(a_i + b_i) = \frac{sum(PM_i)}{|PM_i|}$. Therefore, we have $\sum_{i=1}^{n} CS_i = \sum_{i=1}^{n} \frac{sum(PM_i)}{|PM_i|} = \frac{1}{2}n(n+1)$. □

For illustration, we consider a PM and a C-S vector which appeared in subsection 3.1, that is,

$$PM^{(1)} = \begin{bmatrix} \{1\} \\ \{2,3\} \\ \{2,3\} \\ \{4\} \end{bmatrix}, \text{ and } CS^{(2)} = \begin{bmatrix} 1.5 \\ 1.5 \\ 3 \\ 4 \end{bmatrix}.$$

From Theorem 1, we have

$$\sum_{i=1}^{4} CS^{(2)} = 1.5 + 1.5 + 3 + 4 = 10 = \frac{1}{2}4(4+1)$$

and

$$\sum_{i=1}^{4} \frac{sum(PM_i^{(1)})}{|PM_i^{(1)}|} = \frac{1}{1} + \frac{2+3}{2} + \frac{2+3}{2} + \frac{4}{1} = 10 = \frac{1}{2}4(4+1).$$

Formulas (2), (3) and (4) as well as Figure 1 indicate that the PM and C-S vector are equivalent when they are used for describing ties-permitted ordinal rankings. Theorem 1 highlights a property of the sums of their elements.

## 4. Concluding remarks

The preference map (PM) and Cook-Seiford vector (C-S vector) are both effective concepts as means of describing ties-permitted ordinal rankings in GDM. This paper shows that they are equivalent for representing ties-permitted ordinal rankings. The proposed formulas make it possible to convert mutually between PMs and C-S vectors. Moreover, it is shown that the mutual conversion has the characteristic of inherent consistency.

In GDM, a consensus reaching process aims to improve consensus by evaluating consensus level and identifying disagreements; a selection process aims to produce a final ranking of considered alternatives (Herrera et al., 1996). These two parts constitute a relatively complete GDM procedure. Currently, C-S vectors are used in a traditional distance-based GDM approach which is mainly developed by Cook and Sieford (Armstrong et al., 1982; Cook and Seiford, 1978; Cook, 2006; Cook et al., 1996). The PMs are used in a premetric-based GDM approach (Hou, 2015a; Hou, 2015b; Hou, 2016; Hou and Triantaphyllou, 2018). The premetric-based approach and Cook-Seiford method both have strengths and limitations. The Cook-Sieford approach cannot reflect consensus properly in GDM (Hou and Triantaphyllou, 2018), while it has an efficient selection process as pointed out by Hou (2015a). The premetric-based GDM procedure has an effective consensus reaching process (Hou and Triantaphyllou, 2018). However, the premetric-based procedure also has a limitation of its own, that is, a possible computation complexity in its selection process as a result of the fact that it contains a searching process within a space of $2^{\Omega} - 1$ (Hou, 2015a). Theoretically, this kind of searching process can be an exhaustive job when the cardinality of $\Omega$ is large. Thus a natural way is to take advantage of the strengths of the two methods to form an integrated GDM procedure, meanwhile, with their limitations avoided. The results of this paper paved a possible way for achieving this aim. We will conduct this study in a near future.

**Acknowledgment**: The author was supported by the National Natural Science Foundation of China (No. 71571019).

**References**

Armstrong, R.D., Cook, W.D., & Seiford, L.M. (1982). Priority ranking and consensus formation: The case of ties. *Management Science*, 28(6), 638-645.

Cook, W.D. (2006). Distance-based and ad hoc consensus models in ordinal preference ranking. *European Journal of Operational Research*, 172(2), 369-385.

Cook, W.D., Kress, M., & Seiford, L.M. (1986). An axiomatic approach to distance on partial orderings. *RAIRO-Operations Research*, 20(2), 115-122.

Cook, W.D., Kress, M., & Seiford, L.M. (1996). A general framework for distance-based consensus in ordinal ranking models. *European Journal of Operational Research*, 96(2), 392-397.

Cook, W.D., & Seiford, L.M. (1978). Priority ranking and consensus formation. *Management Science*, 24(16), 1721-1732.

Herrera F, Herrera-Viedma E, Verdegay J.L. (1996). A model of consensus in group decision making under linguistic assessments. *Fuzzy sets and Systems*, 78(1), 73-87.

Hou, F. (2015a). A Consensus gap indicator and its application to group decision making. *Group Decision and Negotiation*, 24(3), 415-428.

Hou, F. (2015b). The prametric-based GDM selection procedure under linguistic as- sessments. fuzzy systems (FUZZ-IEEE). In *Proceedings of the 2015 IEEE interna- tional conference on fuzzy systems* (pp. 1–8). IEEE.

Hou, F. (2016). The prametric-based GDM procedure under fuzzy environment. *Group Decision and Negotiation*, 25(5), 1071-1084.

Hou, F., & Triantaphyllou, E. (2018). An iterative approach for achieving consensus when ranking a finite set of alternatives by a group of experts. *European Journal of Operational Research*, https://10.1016/j.ejor.2018.11.047.